\documentstyle[12pt]{article}
\begin{document}
\renewcommand{\thesection}{\Roman{section}}
\renewcommand{\thetable}{\Roman{table}}
\renewcommand{\thefootnote}{\ast}
\newcommand{\be}{\begin{eqnarray}}
\newcommand{\ee}{\end{eqnarray}}
\newcommand{\zt}{\zeta}
\newcommand{\ve}{\varepsilon}
\newcommand{\al}{\alpha}
\newcommand{\gm}{\gamma}
\newcommand{\Gm}{\Gamma}
\newcommand{\om}{\omega}
\newcommand{\et}{\eta}
\newcommand{\bt}{\beta}
\newcommand{\dt}{\delta}
\newcommand{\Dt}{\Delta}
\newcommand{\la}{\lambda}
\newcommand{\vp}{\varphi}
\newcommand{\nn}{\nonumber}
\newcommand{\nid}{\noindent}
\renewcommand{\baselinestretch}{1.4}
\newcommand{\lmx}[1]{\begin{displaymath} {#1}=
                    \left(\begin{array}{rrr}}
\newcommand{\rmx}{\end{array} \right) \end{displaymath}}

\begin{titlepage}
\rightline{SPbU-IP-99-10}
\vspace{1cm}
\begin{center}
 {\Large \bf Stability of a cubic fixed point in three
 dimensions. \linebreak
 Critical exponents for generic $N$}
\end{center}
\vspace{0.2cm}
\begin{center}
{\bf K. B. Varnashev\footnote{E-mail address: feop@eltech.ru}}
\end{center}
\begin{center}
Department of Physical Electronics, Saint Petersburg Electrotechnical
University, \\
Professor Popov Street  5, St.Petersburg, 197376, Russia
\end{center}

\begin{center}
{\bf  Abstract}
\end{center}
The detailed analysis of the global structure of the renormalization-group
(RG) flow diagram for a model with isotropic and cubic interactions is carried
out in the framework of the massive field theory directly in three dimensions
(3D) within an assumption of isotropic exchange.
Perturbative expansions for RG functions are calculated for arbitrary
$N$ up to the four-loop order and resummed by means of the generalized
Pad$\acute{\rm e}$-Borel-Leroy technique. Coordinates and stability matrix
eigenvalues for the cubic fixed point are found under the optimal value of
the transformation parameter. Critical dimensionality of the model is proved
to be equal to $N_c=2.89 \pm 0.02$ that agrees well with the estimate
obtained on the basis of the five-loop $\ve$-expansion [H. Kleinert and
V. Schulte-Frohlinde, Phys.Lett. B342, 284 (1995)] resummed
by the above method.
As a consequence, the cubic fixed point should be stable in 3D for
$N\ge3$, and the critical exponents controlling phase transitions in
three-dimensional magnets should belong to the cubic universality class.
The critical behavior of the random Ising model being the nontrivial
particular case of the cubic model when $N=0$ is also investigated.
For all physical quantities of interest the most accurate numerical estimates
with their error bounds are obtained.
The results achieved in the work are discussed along with the predictions
given by other theoretical approaches and experimental data.

\vspace{1cm}


\vspace{2cm}
\rightline{{\sl Typeset using} \LaTeX}
\end{titlepage}

\section{Introduction}
\label{sec:1}

At present the critical behavior of the basic models of
phase transitions described by isotropic field theories with a quartic
interaction like $\sum^N_{i=1}(\vp^2_i)^2$ is well studied in the
framework of different theoretical approaches. In particular, the
critical phenomena in polymers, easy-axis ferromagnets, simple liquids,
and binary mixtures, easy-plane ferromagnets, certain superconductors,
as well as superfluid helium-4, Heisenberg ferromagnets, and quak-gluon
plasma in some models of quantum chromodynamics were proved to be
governed by the $O(N)$ symmetric universality class with $N=0,1,2,3$,
and 4, respectively. The large-order field-theoretical renormalization
group (RG) expansions combined with proper resummation techniques,
the high-temperature series method, and the most advanced
Monte Carlo (MC) simulations provided high-precision and comparable
to each other numerical estimates for important physical quantities
(critical exponents, universal couplings, and critical amplitude ratios)
which nowadays are regarded as canonical numbers \cite{Z,GdZ98}.

However in real crystals, due to their complex crystalline structure,
an anisotropy is always present. That is why, when studying phase
transitions in real substances, besides the $O(N)$ symmetric term
one should take into account additional quartic interactions in
corresponding fluctuation Landau-Wilson (LW) Hamiltonian.
The simplest nontrivial crystalline anisotropy is a cubic one.
Proper quartic interaction term is represented as $\sum^N_{i=1} \vp^4_i$.
In this case the vector of magnetization is directed either along the
edges or diagonals of a hypercube in $N$ dimensions of the order parameter
field.

The critical thermodynamics of magnetic and structural phase transitions
in three-dimensional (3D) cubic crystals has been extensively investigated
more than twenty years ago in a good number of papers. By using the
lower-order RG approximations, Wilson and Fisher \cite{WF72}, Aharony
\cite{Ah73}, and Ketley and Wallace \cite{KW73,W73} showed that in the
critical region the fluctuation instability of
continuous phase transitions may be observed, and that it may lead to the
isotropization of the system with a cubic anisotropy. This fact gave rise to
a question what regime of the critical behavior is actually realized in
3D cubic crystal with $N=3$. Much efforts of many people have been devoted
to answer this question. It was understood soon that for the given model
it is enough to calculate the so-called critical (marginal)
dimensionality $N_c$ of the order parameter field. Indeed, the critical
value $N_c$ separates two different regimes of critical behavior of the
system. For $N > N_c$ the cubic rather than the isotropic fixed point
is stable in 3D.
At $N = N_c$ the points interchange their stability so that for
$N < N_c$ the stable fixed point is the isotropic one. Therefore the
calculation of $N_c$ is the crucial point in studying the critical phenomena
in 3D cubic crystal. However attempts to evaluate the critical dimensionality
resulted in the dramatically different estimates.

In fact, the field-theoretical RG analysis of the stability matrix
eigenvalues of the cubic and isotropic fixed points fulfilled
in the one-loop approximation as well as some symmetry arguments
(for details see Sec. III of the paper) lead to the conclusion that
$N_c$ should lie between 2 and 4. Many years ago the three-loop expansion
for $N_c$ as a power series in $\ve$ has been obtained in Ref. \cite{KW73}.
Summation of that short series at $\ve = 1$ (D = 3) by means of the
Pad$\acute{\rm e}$ approximant [1/1] yielded the value $N_c = 3.128$
\cite{Aha}, while making use of the Pad$\acute{\rm e}$-Borel resummation
method results in the estimate $N_c=3$. In contrast
to this, in the work Ref. \cite{YH77}, by using the variational
modification of the Wilson recursion relation method, it has been found that
$N_c = 2.3$. Later, however, Newman and Riedel by decoupling the infinite
system of the recursion relations for the so-called scaling fields and
then solving them showed \cite{NR82} that for $D = 3$ $N_c \sim 3.4$.
At the same time, the classical technique of the high-temperature expansions,
under some circumstances, allowed to establish that for $N = 3$ the isotropic
critical asymptotics in the cubic crystal is unstable \cite{FVdC81},
thus implying that $N_c < 3$.
Further, ten years ago the analysis of the critical behavior of the
(mn)-component field model, which has a good number of interesting
applications to the phase transitions in real substances, has been carried
out within the three-loop RG approach in three-dimensions. The calculation
of the stability matrix eigenvalues for the cubic model ($m = 1$,
$n = 3$) provided the stability of the cubic fixed point in 3D, and the
critical dimensionality turned out to be equal to 2.91 \cite{Sp89}.
In agreement with this, the estimate $N_c = 2.9$ was quoted in the work
Ref. \cite{MSSh89}.
More recently, Kleinert and Schulte-Frohlinde calculated the RG functions
for the cubic model in $(4-\ve)$ dimensions up to the five-loop order
\cite{KS95}.
Resummation of the critical dimensionality expansion with the help of the
Pade approximant [2/2] gave the estimate $N_c = 2.958$ \cite{13}.
The cubic fixed point eigenvalues found by means of a simple resummation
algorithm of the Borel type, accounting the large-order behavior of the
$\bt$-functions when the parameter of anisotropy is very small
\cite{PRd95}, indicated that the cubic point is stable in 3D \cite{KTSf-b}.
Finally, in the recent work Ref. \cite{CH98} by using finite
size scaling techniques and the high-precision MC simulations it has been
suggested that $N_c$ coincides with three exactly. So, such strong scattering
in the estimates of $N_c$ inspired us to study this problem
with particular care. Calculation of the critical dimensionality of the
order parameter field as well as the eigenvalue exponents for the cubic fixed
point by exploiting the higher-order RG approach in three dimensions and
generalized Pad$\acute{\rm e}$-Borel-Leroy (PBL) resummation technique is the
main goal of the paper. As will be shown, our estimates for $N_c$ and
eigenvalues are in excellent agreement with recent results by Kleinert and
collaborators \cite{KTSf-b} obtained on the basis of the five-loop
$\ve$-expansions.

The paper is organized in the following way. In Sec.
II the perturbative expansions for $\bt$-functions of the hypercubic model
are deduced within the RG technique in 3D up to the four-loop order.
In Sec. III the structure of the RG flows of the model is investigated and
fixed point locations for $N\ge 3$ are calculated using the generalized PBL
resummation method.
The eigenvalue exponents of the most intriguing $O(N)$-symmetric and cubic
fixed points are evaluated for the physically important case $N=3$ and their
stability problem is analyzed in Sec. IV. The numerical estimate of the
critical dimensionality $N_c$, at which the topology of the flow diagram
changes, is obtained by resumming both the four-loop RG expansions for the
$\bt$-functions in 3D and the five-loop $\ve$-expansion for $N_c$ at $\ve=1$.
In Sec. V the four-loop RG expansions
of the critical exponents for generic $N$ and their numerical estimates
are presented. Sec. VI is devoted to the study of the critical behavior
of the three-dimensional random Ising model (RIM) which is the special
case of the cubic model when $N=0$. The coordinates, eigenvalues and the
critical exponents of the RIM fixed point are computed by applying the PBL
resummation procedure.
The correction-to-scaling exponent $\om$ is estimated therein.
The results of the investigation are discussed in Conclusion, along with
the predictions and numerical estimates obtained
earlier on the basis of the same or other theoretical approaches and
experimental data.

\section{The model and $\bt$-functions}
\label{sec:2}

The fluctuation LW Hamiltonian of the model reads
\be
H =
\int d^{~3}x \Bigl[{1 \over 2}( m_0^2 \vp_i^2
 + \partial_\mu \vp_i \partial_\mu \vp_i)+
{1 \over 4!} \Bigl(u_0 G_{\>ijkl}^{\>1}+
v_0 G_{\>ijkl}^{\>2}\Bigr)
\vp_i \vp_j \vp_k \vp_l\Bigr] \ , \label{eq:1}
\ee
where $\vp_i$, $i = 1, \ldots, N$ is the real vector
order parameter field in $3D$  and $m^2_0$ is the
linear measure of the temperature, $u_0$ and $v_0$ denote the
"bare" coupling constants. The symmetrized tensors associated with
isotropic and cubic interactions are
\be
 G_{\>ijkl}^{\>1} = \frac{1}{3}\Bigl(
\delta_{ij}\delta_{kl}+
\delta_{ik}\delta_{jl}+
\delta_{il}\delta_{kj}\Bigr) \ , \qquad
 G_{\>ijkl}^{\>2} =
\delta_{ij}
\delta_{ik}
\delta_{il} \ ,
\label{eq:2}
\ee
respectively.
When the cubic symmetry is present, the anisotropic exchange has been
shown to be negligible within the $\ve$-expantion method \cite{B74}.
We assume that the anisotropic exchange is an irrelevant variable in
our case too, and study the critical behavior of model (\ref{eq:1})
with isotropic exchange only.

The model (\ref{eq:1}) has a number of interesting
applications to the phase transitions in simple and complicated
systems. In fact, when $N=1$ Hamiltonian (\ref{eq:1}) describes
the critical phenomena in pure spin systems (the pure Ising model), while
for $N=2$ it corresponds to the anisotropic $XY$ model (the model of
two coupled scalar fields) describing as structural phase transitions
in ferroelectrics as ordering the two-component alloys \cite{WF72,BLgZ74}.
The magnetic and structural phase transitions in a cubic crystal
are governed by model (\ref{eq:1}) as $N=3$. Further, when $N=0$
Hamiltonian (\ref{eq:1}) determines the critical properties
of weakly disordered quenched systems undergoing second-order phase
transitions. The latter is the nontrivial specific case of
the hypercubic model, the systematical studying of which was initiated
in the classical works by Harris and Lubensky \cite{HL74,Lub75} and
Khmelnitskii \cite{Khm76} and then considerably advanced by many authors
when employing the conventional field-theoretical RG approach both in
3D and $(4-\ve)$ dimensions. Finally, the case $N \to \infty$ corresponds
to the Ising model with equilibrium magnetic impurities \cite{Ah73l}.
In this limit the Ising critical exponent of specific heat $\al$ changes
its sign and takes the Fisher renormalization \cite{F68} as well as $\nu$
and $\gm$: $\al \to - \al/(1 - \al)$, $\nu \to \nu/(1 - \al)$,
$\gm \to \gm/(1 - \al)$. Since the critical phenomena
in the pure Ising and anisotropic $XY$ models are well understood,
we will consider below the critical behavior of the cubic and random
Ising models only.

To calculate RG functions the standard normalization conditions of
the massive renormalized theory at fixed dimensions are applied \cite{Par80}.
For each Feynman graph contributing to RG functions the corresponding
contractions are computed by using the algorithm developed in
Ref. \cite{MV98-1}. The combinatorial factors as well as the integral values
are known from Ref. \cite{NMB77}. After performing simple
but cumbersome calculations we obtain the four-loop expansions for the
$\bt$-functions:
\be
 \nn
\bt _u &=& u~ \biggl\{1 - u - {\frac 6{{N+8}}}~ v +{\frac 1{{(N+8)^2}}}
 \biggl[3~ \bigl( 2.024691~ N + 9.382716 \bigr)~ u^2 \\
 \nn
 &+& 44.444444~ u~ v  + 10.222222~ v^2 \biggr]
  - {\frac 1{{(N+8)^3}}} \biggl[3~ \bigl(0.449648~ N^2 \\
 \nn
 &+& 18.313459~ N + 66.546806 \bigr)~ u^3
  + 3~ \bigl(6.646878~ N + 164.613849 \bigr)~ u^2~ v \\
 \nn
 &+& 3~ \bigl(0.621889~ N + 100.955929 \bigr)~ u~ v^2
  + 65.937285~ v^3 \biggr] \\
 \nn
 &+& {\frac 1{{(N+8)^4}}}
 \biggl[- \bigl(0.155646~ N^3 - 35.820204~ N^2 - 602.521231~ N \\
 \nn
 &-& 1832.206732 \bigr)~ u^4 - 3~ \bigl(1.352882~ N^2 - 182.073890~ N \\
 \nn
 &-& 2064.170701 \bigr)~ u^3~ v + 3~ \bigl(27.250336~ N
  + 2110.408809 \bigr)~ u^2~ v^2 \\
 &+& 9~ \bigl(1.291017~ N + 308.599361 \bigr)~ u~ v^3
  + 495.005747~ v^4 \biggr] \biggr\} \ ,
\label{eq:3}
\ee
\be
 \nn
\bt _v &=& v~ \biggl\{1 - {\frac 1{{N+8}}}\bigr(12~ u + 9~ v \bigr)
  +{\frac 1{{(N+8)^2}}}
 \biggl[\bigl( 3.407407~ N + 54.814815 \bigr)~ u^2 \quad \\
 \nn
 &+& 92.444444~ u~ v + 34.222222~ v^2 \biggr]
  - {\frac 1{{(N+8)^3}}} \biggl[- \bigl(1.251107~ N^2 \\
 \nn
 &-& 41.853902~ N - 469.333970 \bigr)~ u^3 + 9~ \bigl(0.248784~ N
  + 136.511768 \bigr)~ u^2~ v \\
 \nn
 &+& 957.781662~ u~ v^2 + 255.929737~ v^3 \biggr]
  + {\frac 1{{(N+8)^4}}} \biggl[\bigl(0.574653~ N^3 \\
 \nn
 &-& 0.267107~ N^2 + 584.287672~ N + 5032.692260 \bigr)~ u^4
  + 3~ \bigl(0.057375~ N^2 \\
 \nn
 &+& 107.641680~ N + 5989.283536 \bigr)~ u^3~ v
  + 3~ \bigl(7321.464604 \\
 &-& 16.494003~ N \bigr)~ u^2~ v^2 + 11856.956858~ u~ v^3
  + 2470.392521~ v^4 \biggr] \biggr\} \ .  \label{eq:4}
\ee

It is reasonable to note that cubic model (\ref{eq:1}) possesses
some specific symmetry property when $N=2$. Namely, the transformation
of the field components
\be
\vp_1 \rightarrow {1\over {\sqrt{2}}}~(\vp_1
+ \vp_2) \ , \qquad
\vp_2 \rightarrow {1\over {\sqrt{2}}}~(\vp_1 -
\vp_2) \label{eq:5}
\ee
combined with substitution of the quartic coupling constants
\be
u \rightarrow u + {3 \over 2}~v \ , \quad v \rightarrow - v
\label{eq:6}
\ee
does not change the structure of the initial Hamiltonian itself.
As a result, $\bt$-functions (\ref{eq:3}) and (\ref{eq:4})
should obey certain symmetry relations \cite{Kor76} which may be
written down as
\be
\bt_u \biggl( u + {3 \over 2}~ v, - v \biggr) &=&
\bt_u ( u, v) + {3 \over 2}~ \bt_v ( u, v) \ ,
\nn \\
\bt_v \biggl( u + {3\over 2}~ v, - v \biggr) &=&
- \bt_v ( u, v) \ . \quad \label{eq:7}
\ee
It can be easily verified that equations (\ref{eq:7}) are really
satisfied. The specific symmetry of Hamiltonian (\ref{eq:1})
will be used below to obtain the lower boundary value of the critical
dimensionality $N_c$.

\section{Method of summation and fixed point locations}
\label{sec:3}

Before analyzing the four-loop $\bt$-functions let us consider briefly the
predictions following from the one-loop approximation. It is easy
to see that $\bt$-functions (\ref{eq:3}) and (\ref{eq:4}) in this
approximation have four different solutions corresponding to Gaussian
(trivial), Ising, isotropic (Heisenberg) and cubic fixed points with the
coordinates:
\newpage
\be
&1)& u^G_c = v^G_c = 0 \ ,
\nn \\
&2)& u^I_c = 0 \ , \quad v^I_c = {(N + 8) \over 9} \ ,
\nn \\
&3)& u^H_c = 1 \ , \quad v^H_c = 0 \ ,
\nn \\
&4)& u^C_c = {N + 8 \over {3 N}} \ , \quad
v^C_c = {(N - 4) (N + 8) \over {9 N}} \ ,
\qquad \qquad \qquad \qquad \qquad \qquad \qquad \qquad
\label{eq:8}
\ee
respectively. The most intriguing fixed points are the isotropic and
cubic ones. At $N = 4$ the coordinates of these two points coincide that
leads to the conclusion that the critical dimensionality $N_c$ has the upper
boundary value $N_c = 4$. On the other hand, the eigenvalue exponents
for the isotropic and cubic fixed points are given by the expressions:
\be
\la^H_1 &=& -1 \ , \qquad \la^H_2 = (N - 4)~/~(N + 8) \ ,
\nn \\
\la^C_1 &=& -1 \ , \qquad \la^C_2 = (4 - N)~/~3N  \ . \qquad
\label{eq:9}
\ee
If the real parts of both eigenvalues are negative, the corresponding
fixed point is infrared stable; if eigenvalues are of opposite sings,
the point is of "saddle-knot" type. It is seen from Eqs.(\ref{eq:9})
that $N_c = 4$ actually separates two different regimes of critical
behavior of the model. When $N > N_c$ the cubic rather than the isotropic
fixed point is stable in 3D, while for $N < N_c$ the stable fixed point
is the isotropic one.

To determine the lower boundary of $N_c$ one should employ the above
mentioned specific symmetry property of the model, when $N=2$ \cite{MS87}.
As was already pointed out, the rotation (\ref{eq:5}) of the components of
$\vp_i$ by $\pi/4$ combined with substitution (\ref{eq:6}) generates
relations (\ref{eq:7}), but does not
change the form of the RG equations. However for $N=2$ transformations
(\ref{eq:5}), (\ref{eq:6}) result in the relocation of the coupling
constants values
so that the cubic and Ising fixed points are transformed into each another
at the 3D RG flow diagram. Since the exact RG equations always have the
Ising fixed point, which inevitably is the saddle-knot one, these equations
should have also the cubic fixed point, which will be unstable. In this
situation, the isotropic fixed point, again always existing in the exact
RG equations, should be the stable knot only. Therefore we conclude that the
lower boundary of $N_c$ is not less than two. Of course, the
real value of $N_c$ can be obtained only on the basis of the thorough
analysis of the structure of the RG flow diagram, provided that the
$\bt$-functions of the model are calculated in sufficiently high-order RG
approximations and then processed by means of appropriate resummation
techniques.

Let us now concentrate our attention on the analysis of the four-loop
$\bt$-functions. It is well known that field-theoretical RG expansions
are divergent. The character of their large-order asymptotic behavior
for the case of simple $O(N)$-symmetric models was established in
Refs. \cite{Lip77,BLgZ77,BrPr78}. In particular, it was proved that
the coefficients of the series at large $k$ behave as
$c (-a)^k k! k^b$, where the asymptotic parameters $a$, $b$ and $c$
are assumed to be calculated for each RG function. Knowledge of the exact
values of the asymptotic parameters in combination with the most powerful
resummation procedure of the Borel transformation with a conformal
mapping \cite{Loef76}, first proposed in Ref. \cite{LgZ77} and then
elaborated in Refs. \cite{SZ79,KTSh79,VKT79}, made it possible to develop
the accomplished quantitative theory of critical behavior of simple systems
\cite{Z,GdZ98,LgZ77,VKT79,LgZ80,GLT84,LgZ85}.

At the same time, the asymptotic nature of RG functions of anisotropic
models is unknown. Calculation of the exact values of the
asymptotic parameters characterizing the large-order behavior of the series
in such models is a very difficult and still unsolved problem. As an exception
one should mention the anisotropic quartic quantum oscillator representing
one-dimensional $\vp^4$ field theory with a cubic anisotropy.
Within an assumption of the weak anisotropy, the transformation parameters
for the perturbative expansion of the ground state energy of this system
as well as for the $\bt$-functions of the cubic model have recently been
found \cite{PRd95,KTJ-a}. Later this information has been used to solve
the stability problem of the cubic fixed point in three dimensions
from $\ve$-expansion \cite{KTSf-b}.

Usually, in a lack of any information about the high-order behavior
of the series either the simple Pad$\acute{\rm e}$-Borel or Chisholm-Borel
resummation procedures are used, for treating the perturbative expansions
of anisotropic models. The latter technique, however, possesses at least
two inherent drawbacks.
First, some ambiguity in the calculation of coefficients
of denominators of the Chisholm approximants is unavoidable \cite{Ch73}.
Second, the Chisholm-Borel procedure does not hold the specific symmetry
properties of a model.
At the same time, exploiting the Borel transformation in combination with
the Pad$\acute{\rm e}$ or Chisholm approximants shows that the results of
calculation are very sensitive to the choice of the type of approximants.
This may lead to the estimates, which do not provide reliable predictions
even in the higher-loop RG approximations (see, for instance, Sec. VI of
the paper).
Besides, in the framework of both schemes it is very difficult to determine
any error bounds for evaluated quantities.

In the present work we apply for processing RG expansions of the
$\bt$-functions and critical exponents of model (\ref{eq:1}) the PBL
resummation method generalized for the two coupling constant case.
This resummation technique, first introduced by Baker, Nickel,
and Meiron in Ref. \cite{BNM78}, turned out to be highly efficient when used
to study the critical behavior of simple $O(N)$-symmetric models in 3D.
The critical exponent estimates obtained within the framework of this
technique are regarded nowadays as the most accurate values, as those of
Ref. \cite{GdZ98,LgZ77,LgZ80}. We motivate our choice of the PBL resummation
method by the following reasons:
\begin{itemize}
\item 3D RG expansions for the $\bt$-functions and critical exponents
      of the cubic model are alternating in signs. Therefore using the
      PBL resummation technique is quite natural.
\item It can be expected that for complex models with more than one
      coupling constant, the asymptotics of RG series at large orders
      will comprise a factor $k! k^b$. The PBL resummation method removes
      divergences of this type.
\item The PBL resummation method allows one to determine the error bounds
      for the physical quantities to be calculated, in a natural way.
\end{itemize}

\nid
The generalized PBL resummation procedure consists of the following steps.
Let a physical quantity $F(u,v)$ be represented by a double series
\be
F(u, v) = \sum_{i,j} f_{ij} u^i v^j \ ,
\label{eq:10}
\ee
where coefficients $f_{ij} \sim (i+j)! (i+j)^b$ at large orders
$(i,j \to \infty)$, the additional  parameter $b$ being an arbitrary
non-negative number to be defined below. Associated with the initial
series (\ref{eq:10}) is the function
\be
{\cal F}(u, v; b) = \int\limits_0^\infty
e^{-t} t^b B(ut, vt) dt \ \
\label{eq:11}
\ee
The Borel-Leroy transform $B(x,y)$ is the analytical
continuation of its Taylor series
\be
B(x, y) = \sum_{ij} {f_{ij} \over {\Gm(i+j+b+1)}} x^i y^j
\label{eq:12}
\ee
absolutely convergent in a circle of the nonzero radius. In order to
calculate the integral in (\ref{eq:11}) one should continue analytically
$B(x,y)$ for $0 \le x < \infty$ and $0 \le y < \infty$.
To this end, the rational Pad$\acute{\rm e}$ approximants [L/M] $(x,y)$
are used. The Pad$\acute{\rm e}$ approximant method is determined in
a conventional way \cite{BGm}. Let us consider a "resolvent" series
\be
\tilde B (x, y, \la) = \sum_{k = 0}^{\infty} \la^k
\sum_{l = 0}^k {f_{l, k-l} x^l y^{k-l}
\over {\Gm(k+b+1)}} = \sum_{k = 0}^{\infty} A_k \la^k \ ,
\qquad \qquad
\label{eq:13}
\ee
where coefficients $A_k$ are uniform polynomials of $k$th order in $u$
and $v$. The sum of the series is then approximated by
\be
B (x, y) = [ L / M ] \Big \arrowvert_{\la = 1} \ .
\label{eq:14}
\ee
The Pad$\acute{\rm e}$ approximants [L/M] in $\la$ are given by an
attitude
\be
[L/M] = {P_L (\la) \over {Q_M (\la)}} \ , \label{eq:15}
\ee
where $P_L (\la)$ and $Q_M (\la)$ are polynomials of degrees $L$
and $M$ respectively with coefficients depending on $x$ and $y$ which
should be determined from the conditions:
\be
&Q_M& (\la) \tilde B (x, y; \la) - P_L (\la) =
O (\la^{L + M + 1}) \ , \nn \\
&Q_M& (0) = 1 \ .
\label{eq:16}
\ee
Replacing variables $x = u t$ and $y = v t$ in the Pad$\acute{\rm e}$
approximants and then evaluating the Borel-Leroy integral
\be
{\cal F} (u, v; b) = \int\limits_0^{\infty} e^{-t} t^b [L/M]
\Big \arrowvert_{\la = 1} dt \ .
\label{eq:17}
\ee
we obtain the approximate expressions for RG functions.

Among the Pad$\acute{\rm e}$ approximants the diagonal ($L = M$) or
near-diagonal ones were proved to exhibit the best approximating
properties \cite{BGm}.
However, as the degree of the denominator $M$ increases, the number of
possible poles of the approximant increases too. If some of the
poles belong to the positive real semiaxis, the
corresponding approximant should be rejected. Due to this, the choice of
"working" approximants, which might be used for analytical continuation
of the Borel-Leroy image onto the complex cut plane, is largely limited.
On the other hand, varying the free parameter $b$ in Borel-Leroy
transformation (\ref{eq:11}) allows one to optimize the resummation
procedure under the condition that the fastest convergence of the iteration
process is achieved. So, taking into account the above mentioned remarks,
in order to find
the fixed points locations of the model we adopt the following scheme. For
the fixed $N$, the $\bt$-functions are resummed by virtue of transformation
(\ref{eq:11}) in the highest-loop orders by shifting the transformation
parameter $b$. In order to make an analytical continuation of the
Borel-Leroy transforms $B_u (u,v)$, $B_v (u,v)$ over the cut-plain, the most
appropriate Pad$\acute{\rm e}$ approximants [2/1], [3/1], and [2/2]
are chosen.
The fixed points locations are then determined for each $b$ from solution
of a set of equations: $\bt^{res}_u (u_c, v_c) = 0$,
$\bt^{res}_v (u_c, v_c) = 0$. The final locations are obtained
by averaging over the values given by the approximants under the optimal
value of the parameter $b$, at which the quantity
$|1 - {\cal F}_L (u,v;b)/{\cal F}_{L-1} (u,v;b)|$ reaches its local minima.
The quantity ${\cal F}_L (u,v;b)$ is evaluated
for the $L$-partial sum of the series in Eq.(\ref{eq:17}), $L$ stands for
the step of truncation of the series.

In Fig. 1 the results of computation of the cubic fixed point locations
depending on the parameter $b$ are presented for the physical important case
$N = 3$. Three curves correspond to the three Pad$\acute{\rm e}$
approximants. The parameter $b$ shifts from $0$ to $3$. As seen from
the figure the optimal value of $b$ is zero. At this point the
numerical values of the cubic fixed point locations given by different
approximants are the most close to each other.
The result of computing the cubic fixed point locations are also presented
in Table I. In the first three columns of the table the fixed
point locations values found for the Pad$\acute{\rm e}$ approximants
[2/1], [3/1], and [2/2] at $b=0$ are placed.
Averaging the results of processing over the all approximants under the optimal
value of $b$ gives the estimates standed in the fourth column of the table.
These numbers we adopt as the final estimates of the cubic fixed point
locations found within the four-loop approximation.
As an accuracy for these approximate values we take the maximum deviations
of the average values of the fixed point locations from those given by the
approximants at $b=0$.

One can observe, looking at Fig. 1,
that the values of the cubic fixed point locations given by the symmetric
approximant [2/2] weakly depend on the shift parameter $b$. Averaging over the
all values given by this approximant within the interval [0,3] results in
the cubic fixed point locations estimates presented in the fifth column
of Table I.
The coordinates of the cubic fixed point found earlier on the basis of the
three- and four-loop approximations with the use of the Chisholm-Borel
resummation method are presented in the sixth and seventh columns of the
table for comparison. These numbers include the normalizing multiplier
$11 \over 9$ needed to compare our $\bt$-functions with those obtained
in Refs. \cite{Sp89,MSSh89}.

In order to verify the correctness of the chosen approach let us apply
the above considered scheme to estimate the fixed point locations of the
$O(N)$-symmetric model where the numerical results are well known.
Consider, for example, an $O(3)$-symmetric case relevant to the
Heisenberg ferromagnets.
The six-loop 3D RG expansion for the $\bt$-function of this model was
reported in Refs. \cite{LgZ77,BNM78}. The PBL resummation of that
series with the use of eight types of the Pad$\acute{\rm e}$
approximants [2/1], [3/1], [2/2], [4/1], [3/2], [5/1], [4/2], and [3/3]
yields, after solving the equation $\bt^{res} (g_c) = 0$, the picture displayed
in Fig. 2. It is seen that the values of the isotropic fixed point location
calculated in the highest RG orders with the help of the approximants [3/3],
[4/2], and [3/2] are very weakly dependent on the parameter $b$ varied
within the interval $0 \le b \le 15$.
The curves corresponding to these approximants are intersected at $b=4.5$.
Therefore the value $b=4.5$ is the optimal value of the transformation
parameter in which the fastest convergence of the iteration procedure
is ensured. For $b=4.5$ the central value estimate
of the isotropic fixed point is $g_c = 1.392$. The maximum deviation of
the central value from the values given by some of the approximants [3/3],
[4/2], and [3/2] at the point $b=10$ is adopted approximately as an apparent
accuracy of the calculation, $\Dt = 0.0013$. Such a small error can be
explained by the small dispersion of the curves within
the range $5 \le b \le 10$. So, the estimate $g_c = 1.3920 \pm 0.0013$
is in excellent agreement with those found more then twenty years ago in
Refs. \cite{LgZ77,BNM78} as well as with recent results of Ref. \cite{GdZ98}.

Within the framework of the four-loop approximation there are only three
appropriate Pad$\acute{\rm e}$ approximants.
Averaging the results of computing the isotropic fixed point location
given by the approximants [2/1], [3/1], and [2/2] under the optimal value
of the transformation parameter results in the estimate
$g_c = 1.3925 \pm 0.0070$. The error was determined again through the maximum
deviation of the central value from those given by each of the approximants
at $b=0$. It is seen that the four-loop estimate of the coordinate of the
isotropic fixed point is in a good accordance with the best ones followed
from the six-loop consideration.

Let us note that the coordinate of the $O(3)$-symmetric fixed point
calculated within the five-loop approximation does not approach the exact
value. Namely, the PBL resummation procedure leads to the estimate
$g_c = 1.3947 \pm 0.0040$. Although the error of the calculation became
visibly smaller, the central value of the fixed point location stepped
aside from the four- and six-loop ones.

Thus, the fulfilled numerical
analysis shows that the isotropic fixed point location estimate obtained
in the four-loop level occurs to be close to the "exact" value.
Therefore it can be expected that in the case of the cubic model the
fixed point locations $u_c=1.3428 \pm 0.0200$, $v_c=0.0815 \pm 0.0300$
(see fourth column of Table I) will not distinguish strongly from "exact"
values as well. The coordinates of the cubic fixed point for some values
$N$ of the order parameter dimensionality are presented in Table II.
Our calculations show that for $N=3$ the coordinates of the cubic fixed point
practically do not differ from those of the Heisenberg one. However, with
increasing $N$ the cubic fixed point runs away from the isotropic
point moving towards the Ising one. In the large $N$ limit these two fixed
points become close to each another so much that the influence of the
$O(N)$-symmetric invariant on the critical thermodynamics of the cubic model
vanishes. This can be easily seen by applying the
$\frac{1}{N}$ consideration to the one-loop solutions of the RG equations
of model (\ref{eq:1}). Indeed, rescaling the coupling constants
$u \rightarrow u/N$, $v \rightarrow v/N$ in the initial Hamiltonian and
taking then the limit $N \to \infty$ in Eqs.(\ref{eq:8}) one can see that
the cubic fixed point approaches the Ising one asymptotically.
So, the cubic model turns out to be splitted into $N$ non-interacting Ising
models, the critical behavior of each of them is determined by a set
of critical exponents renormalized according to Fisher \cite{F68}.

The data listed in Table II will be used further for calculating the
stability matrix eigenvalues as well as the critical exponents of the
hypercubic fixed point.

\section{Stability and critical dimensionality}
\label{sec:4}

One of the independent way to determine fixed point locations in fixed $D$
is to construct the RG flow phase diagram of a model.
If, at the flows diagram, there exists a fixed point of stable knot type,
the trajectories originated from some point within the range of stability
of the initial Hamiltonian would flow towards the knot.
The region at the flow diagram
where the trajectories are intersected provides the coordinates of
this stable fixed point. Thoroughly investigating the 3D RG flow
diagram of cubic model (\ref{eq:1}) in the four-loop approximation
we arrived at the conclusion that the cubic rather than the isotropic fixed
point is absolutely stable for all $N \ge 3$.

On the other hand, the reliable conclusion about the stability of the
cubic fixed point for $N \ge 3$ can be given on the basis of calculating
the eigenvalue exponents $\la$'s of the stability matrix
\lmx{M_{ij}}
    \frac{\partial\bt_u}{\partial u} &
    \frac{\partial\bt_u}{\partial v} \\
    \frac{\partial\bt_v}{\partial u} &
    \frac{\partial\bt_v}{\partial v}
\rmx
taken at $u=u_c$ and $v=v_c$. If the real parts of both eigenvalues are
negative, the fixed point is the stable knot on the $(u,v)$ plane.
If $\la_1$, $\la_2$ have opposite signs, the point is of
the "saddle--knot" type.

To calculate the stability matrix eigenvalues of the cubic and isotropic
fixed points we have chosen the following strategy. First, the derivatives of
the $\bt$-functions (\ref{eq:3}), (\ref{eq:4}) are calculated, and new RG
expansions resummed by means of the PBL technique are substituted into
the matrix $M_{ij}$. The eigenvalue exponents of the matrix of derivatives
$M_{ij}$ obtained in this way are evaluated then under the optimal value
of the transformation parameter $b$. In Fig. 3 a), b) we present our
numerical results for the series $-\frac{\partial\bt_u}{\partial u}$ and
$-\frac{\partial\bt_v}{\partial v}$ for the physically interesting
case $N=3$. The curves correspond to the three types of the
Pad$\acute{\rm e}$ approximants used within the four-loop approximation.
The crossing of the curves gives the optimal value of $b$ at which
we find $\frac{\partial\bt_u}{\partial u}|_{opt} = -0.7536$ and
$\frac{\partial\bt_v}{\partial v}|_{opt} = -0.0331$. Because the series
$-\frac{\partial\bt_u}{\partial v}$ and $-\frac{\partial\bt_v}{\partial u}$
are turned out to be shorter by one order in comparison with
$-\frac{\partial\bt_u}{\partial u}$ and $-\frac{\partial\bt_v}{\partial v}$,
their resumming performed with the help of the approximant [2/1] only yields
the monotonic dependence of the result of processing on the parameter $b$.
In this unfavorable situation, we take into account an additional
Pad$\acute{\rm e}$ approximant [1/1] to optimize the iteration procedure.
The results are plotted in Fig. 3 c), d). For the optimal values of
$b$ we obtain $\frac{\partial\bt_u}{\partial v}|_{opt} = -0.4566$
and $\frac{\partial\bt_v}{\partial u}|_{opt} = -0.0409$. Straightforward
calculation of the eigenvalues of the stability matrix $M_{ij}$ gives for
the cubic fixed point the numbers placed in Table III.
The eigenvalues of the isotropic fixed point as well as the analogous
numerical estimates obtained recently in Ref. \cite{KTSf-b} on the basis of
using the five-loop $\ve$-expansions are presented for comparison therein.
These estimates show that the cubic fixed point is absolutely stable in 3D
for $N=3$, while the isotropic fixed point appears to be stable on the
$u$-axis only. Our numerical results agree well with those obtained
in Ref. \cite{KTSf-b}.

Let us now calculate the critical dimensionality $N_c$ of the order parameter
field. The critical dimensionality is defined as a value of $N$ at
which the cubic fixed point coincides with the isotropic one. Equivalently,
for $N=N_c$ the second eigenvalue of the stability matrix $M_{ij}$ vanishes,
$\la_2 = 0$.

Studying carefully the 3D RG flow diagram of model (\ref{eq:1}) depending
on the order of approximation
with the use of different Pad$\acute{\rm e}$ approximants we
arrive at the conclusion that $N_c=2.910 \pm 0.035$ and $N_c=2.890 \pm 0.020$
within the three- and four-loop approximations respectively. The accuracy of
calculation of $N_c$ was determined through the evaluation of the stability
matrix eigenvalues for different $N$ from the interval of the above
mentioned errors.
That value of $N=N_c$, above or below of its central number, at which the
second eigenvalue $\la_2$ was becoming nonzero, was taking for the upper or
lower boundary of $N_c$ respectively.

It is worthy to compare the four-loop estimate of $N_c$ just found with that
which can be obtained within the $\ve$-expansion method. The five-loop
$\ve$-expansion for $N_c$ has been calculated in Ref. \cite{KS95}.
The series turned out to be alternating in signs that allows one to resum
it by means of the PBL technique. To this end, we will use again
the most appropriate
Pad$\acute{\rm e}$ approximants [2/1], [3/1] and [2/2] for analytical
continuation of the Borel-Leroy transform for all $0 \leq \ve t \le \infty$.
Dependence of the results of processing the critical
dimensionality $N_c$ on the transformation parameter $b$ is depicted in
Fig. 4. The curves corresponding to the approximants are crossed at
the point $b \sim 1$. The appropriate value of the critical dimensionality
is $N_c=2.894 \pm 0.040$. As an error of the calculation it is natural
to assume the maximum scattering of numerical values given by the
approximants at $b=0$ with respect to the value obtained at the
crossing point of the curves. This estimate of $N_c$
is in excellent agreement with the above found within the 3D RG approach.
So, both schemes, the RG technique directly in 3D and the $\ve$-expansion
method, result in the same estimate of the critical dimensionality $N_c=2.89$,
thus implying that the cubic fixed point is stable in three dimensions for
$N \ge 3$. This means that the critical behavior of model (\ref{eq:1})
should be governed by the cubic fixed point with a certain set of critical
exponents which will be calculated for generic $N$ in the next section.

\section{Critical exponents for generic $N$}
\label{sec:5}

Having the coordinates of the cubic fixed point, the stability of which
in 3D for $N \ge 3$ has been proved in the previous section, it is possible
to obtain the numerical estimates for the critical exponents.
Two of them are known to determine the critical behavior of the system while
the others can be found via the famous scaling laws. The four-loop RG
expressions for the magnetic susceptibility exponent $\gm^{-1}$ as well as
for the correlation function exponent $\et$ for generic symmetry index
$N$ are as follows:
\be
 \nn
\gm^{-1} &=& 1 - {\frac 1{{N+8}}} \biggl[{\frac {(N+2)}{2}}~ u
  + {\frac 3{2}}~ v \biggr] + {\frac 1{{(N+8)^2}}}
 \biggl[(N+2)~ u^2 + 6~ u~ v + 3~ v^2 \biggr] \\
 \nn
 &-& {\frac 1{{(N+8)^3}}} \biggl[ \bigl(0.879559~ N^2 + 6.485477~ N
  + 9.452718 \bigr)~ u^3 + 9~ \bigl(0.879559~ N \\
\quad \label{eq:18}
 &+& 4.726359 \bigr)~ u^2~ v + 9~ \bigl(0.128340~ N
  + 5.477578 \bigr)~ u~ v^2 + 16.817754~ v^3 \biggr] \\
 \nn
 &+& {\frac 1{{(N+8)^4}}}
 \biggl[- \bigl(0.128332~ N^3 - 7.966741~ N^2 - 51.844213~ N
  - 70.794806 \bigr)~ u^4 \\
 \nn
 &-& 3~ \bigl(0.513328~ N^2 - 32.893620~ N - 141.589613 \bigr)~ u^3~ v
  + 9~ \bigl(3.423908~ N \\
 \nn
 &+& 83.561044 \bigr)~ u^2~ v^2+ 27~ \bigl(0.208999~ N
  + 19.120991 \bigr)~ u~ v^3 + 130.477428~ v^4 \biggr] ,
\ee
\be
 \nn
\et &=& {\frac 1{{(N+8)^2}}}
 \biggl[0.296296~ (N+2)~ u^2 + 1.77777~ u~ v + 0.888888~ v^2 \biggr] \\
\quad \label{eq:19}
 &+& {\frac 1{{(N+8)^3}}} \biggl[0.024684~ (N+2)~ (N+8)~ u^3
  + 0.222156~ (N+8)~ u^2~ v \\
 \nn
 &+& 1.99940~ u~ v^2 + 0.666468~ v^3 \biggr] + {\frac 1{{(N+8)^4}}}
 \biggl[-\bigl(0.004299~ N^3 - 0.667985~ N^2 \\
 \nn
 &-& 4.60922~ N - 6.51210 \bigr)~ u^4 - 3~ \bigl(0.017194~ N^2 - 2.70633~ N
  - 13.0242~ \bigr)~ u^3~ v \\
 \nn
 &+& 3~ \bigl(0.681615~ N + 22.8884 \bigr)~ u^2~ v^2
  + 47.140~ u~ v^3 + 11.7850~ v^4
 \biggr] .  \\ \nn
\ee
These series, however, are known to be divergent.
To extract from them a physical information concerning the critical behavior
of the substances of interest
we will apply the same resummation procedure as that used above.
Note, since the coefficients of the series of $\et$ are rapidly diminishing,
its values are found by the direct substitution of the coordinates of the cubic
fixed point into Eq.(\ref{eq:19}), whereas the series of the exponent
$\gm^{-1}$ before the substitution needs to be resummed.
Using the generalized PBL resummation technique, we
evaluate the magnetic susceptibility exponent $\gm$ and then, having
the values of $\et$ at hand, estimate the correlation length critical exponent
$\nu$ by the scaling relation. The result of numerical processing
Eq.(\ref{eq:18}) depending on the parameter $b$ for $N=3$ is dipicted
in Fig. 5. The three types of Pad$\acute{\rm e}$ approximants have
been used, and the transformation parameter shifted within the range
[0,15]. For the optimal value of $b$ we obtain
$\gm = 1.3775 \pm 0.0040$. Taking into account that in the four-loop
approximation $\et = 0.0332 \pm 0.0030$, we arrive at the estimate of the
critical exponent $\nu$: $\nu = 0.6996 \pm 0.0037$. The error bounds
for the exponents $\gm$ and $\nu$ have been determined through a maximum
deviation of their central values at $b=b_{opt}$ from the corresponding
values given by the approximants at $b=0$.
In the case of the exponent $\et$, as the uncertainty we take the absolute
difference between two successive, three- and four-loop,
results. The critical exponents estimates of the cubic fixed point for other
values of $N$ are listed in Table II. The estimates obtained from the
$\ve$-expansions \cite{MV98e} are placed for comparison therein.

How much do the critical exponent estimates found differ from the
"exact" ones? To answer this we resort again to the analysis of numerical
results for the $O(3)$-symmetric model obtained on the basis of using
the PBL procedure.
In Fig. 6 the results of processing the 3D RG series for the susceptibility
exponent $\gm$ in successive orders of perturbation theory in a number
of loops are presented. Within the four-loop approximation under the optimal
value of transformation parameter we find $\gm = 1.3778$.
The six-loop calculations under the optimal choice of $b$ (the point of
crossing the curves corresponding to the approximants [4/2], [5/1],
and [4/1]) give $\gm = 1.3867$.
The latter estimate seems to be in a good agreement with that
known from Refs. \cite{LgZ77,BNM78}. Thus, the difference
between the four-loop results and their six-loop counterparts
does not exceed 0.01.

For $N=2$ the numerical results for the cubic model turn out to be much
better.
In this case the cubic fixed point lies in the $v < 0$ half plane and is
unstable ($N < N_c$). However for $N=2$ the Hamiltonian of model
(\ref{eq:1}) possesses the specific symmetry [see Eqs.(\ref{eq:5}),
(\ref{eq:6})] that transforms the cubic point into the Ising one and vice
versa. Due to the symmetry, the critical exponents of both fixed points
coincide. In the four-loop level for the cubic fixed point at $N=2$
we found the estimates:
$\gm=1.2416$, $\eta=0.0323$, and $\nu=0.631$.
These numbers agree well with the best estimates followed from
the six-loop RG expansions \cite{GdZ98,LgZ77,LgZ80,BNM78}.
One may hope therefore that the critical exponent estimates for the cubic
model obtained in the present work within the four-loop approximation will
differ from the "exact" values by no more than 1--2\%.

Unfortunately, for the most important case $N=3$ having a good number
of interesting applications to the critical phenomena in real substances,
the critical exponents of the cubic and isotropic fixed points turn out to be
practically the same. This is the consequence of the closeness of both
points on the 3D RG flow diagram. Thus, although in the course of the
investigation
the cubic fixed point has been shown to be stable at $N=3$ and, therefore,
the critical behavior of the magnetic phase transitions in crystals with
cubic anisotropy should belong to the cubic rather than the isotropic
universality class, a certain difficulty arises in trying to identify the
cubic fixed point from experimentally determined exponents. Due to this
"near-marginality", the calculation of the critical exponents in cubic magnets
seems to be of academic interest only.

At the same time, the numerical analysis shows that as $N$ increases the
distinction between the critical exponents of the cubic and isotropic fixed
points increases as well. In the limit $N \to \infty$ the critical exponents
of the cubic fixed point go over into those of the Ising model with
equilibrium magnetic impurities \cite{F68}.

\section{Numerical results for the RIM}
\label{sec:6}

In the critical region the character of critical behavior of the defect
crystals such as impure uniaxial ferromagnets $Li Tb_{1-x} Y_{x} F_4$,
$Cd_{1-x} Nd_{x} Cl_3$ or diluted Ising antiferromagnets
$Mn_x Zn_{1-x} F_2$, $Fe_x Zn_{1-x} F_2$ are known to be described
by the 3D random Ising model with the effective Hamiltonian
\be
H =
\int d {\bf x} \Bigl[{1 \over 2}( m_0^2 \vp^2
 + (\nabla \vp)^2) +
{1 \over 4!} v_0 \vp^4 + \psi({\bf x}) \vp^2 \Bigr] \ ,
\label{eq:20}
\ee
where $\psi({\bf x})$ is the static random field describing fluctuations
of local
transition temperature, $m_0^2 - m_{0_c}^2 \sim T - T_c$. Averaging over all
configurations of $\psi({\bf x})$ with Gaussian weight and employing the
replica trick \cite{Zieg} one can reduce problem (\ref{eq:20})
to the analysis of critical behavior of the $N$-component hypercubic model
(\ref{eq:1}) in the limit $N \to 0$ \cite{GrLutAh}. Moreover, the Ising
vertex $v_0$ in Eq.(\ref{eq:20}) plays a role of the cubic vertex in
Eq.(\ref{eq:1}), while a role of the impure vertex plays the isotropic one
$u_0$. Obviously, because now $u_0 < 0, v_0 > 0$, the added interaction of
critical fluctuations of the order parameter field through the
intermediary of impurities is the attraction.

Studying the magnetic and structural phase transitions in weakly
disordered systems is of considerable interest both from
theoretical and experimental point of view. It is well known that the
critical exponents of such systems should differ markedly from those
of the pure ones, due to the famous Harris criterion \cite{HarCr}.
Indeed, according to Harris, provided the specific-heat exponent
$\al_{\sl pure}$
of a pure system is positive, i.e. the specific-heat is divergent at the
critical point ($C \sim |T - T_c|^{-\al}$), a new critical behavior under
dilution is expected.
Otherwise, if $\al_{\sl pure} < 0$, the critical behavior of the random
system would be similar to that of the pure one.
Among the three-dimensional $N$-vector models, only the Ising model has
$\al > 0$, and the corresponding new critical behavior has been obtained
as an unusual stable fixed point dependent upon $\sqrt{\ve}$, where
$\ve = 4 - D$. Furthermore, one (smallest) of the eigenvalues of the
stability matrix taken at the RIM fixed point is expressed through
the critical exponents $\al$ and $\nu$: $\la_2 \sim \al / \nu$
(for the details see \cite{Sak,SkMa}).

The systematical calculation of the RIM critical exponents was
hystorically begun in the seminal works by Harris and Lubensky
\cite{HL74,Lub75} and Khmelnitskii \cite{Khm76}. However the $\ve$-expansion
technique could not provide the reliable numerical estimates, because
RG equations of model (\ref{eq:1}) for $N=0$ turn out to be
degenerate in the one-loop approximation. Such a degeneracy causes powers
of $\sqrt{\ve}$ to appear in expansions for the fixed point locations as well
as for the critical exponents, thus leading to the substantial decrease of
accuracy expected within the high-loop approximations \cite{ShAS,MV98-2}.

The following pronounced step to evaluate the RIM
exponents was made by G.Jug \cite{Jug}, who applied the alternative
approach, the RG in fixed dimensions. The reasonable numerical
estimates were obtained within the two-loop approximation by making use of
the Chisholm-Borel procedure to resum the resulting series. Later the
critical exponents series for the 3D RIM were deduced within the three-
and four-loop approximations and corresponding numerical estimates were
obtained on the basis of the Chisholm-Borel summation method
\cite{Sp89,MSSh89}, Pad$\acute{\rm e}$-Borel procedure and the first
confluent form of the $\ve$ algorithm of Wynn \cite{M89}.

In this section we will study the critical thermodynamics of the 3D RIM
using the generalized PBL resummation method. Setting $N=0$
in Eqs.~(\ref{eq:3}) and (\ref{eq:4}) and solving the system of equations
$\bt^{res}_u (u_c,v_c)=0$, $\bt^{res}_v (u_c,v_c)=0$ we find the RIM
fixed point locations depending on the transformation parameter $b$. Resulting
curves are depicted in Fig. 7. Unfortunately, this picture is not complete,
because in the three-loop approximation the Pad$\acute{\rm e}$ approximants
[2/1] and [1/2] have the poles for all $b$. On the other hand, the fixed point
locations values given by the approximant [1/1] seem to be very far from
the true ones. So, in order to determine the RIM fixed point locations
we cannot apply the optimization algorithm described in Sec. III, at least
within the given approximation. In such situation we need a new working
criterion. We can, for instance, select the approximants providing
the most stable values under the variation of the parameter $b$.
As is seen from Fig. 7 the locations of the RIM fixed point
given by the approximants [2/2], for the $u_c$-component, and [3/1], for the
$v_c$-component, are practically independent on the parameter $b$. Indeed, the
dispersion of the corresponding curves within the range $0 \le b \le 4$ is
no more than $1 \cdot 10^{-4}$.
The fixed point locations obtained in such a way may then be used
for calculation of the critical exponents $\gm$, $\eta$, and $\nu$
as well as of the stability matrix eigenvalues when starting the
optimization procedure (see Secs. IV and V). Corresponding numerical
estimates are summarized in Table IV. For comparison we collected in the table
the data obtained earlier either by resumming RG functions
within the minimal subtraction scheme directly at $D=3$ (3D MS) \cite{FHY98},
or by applying different resummation procedures (Chisholm-Borel
technique \cite{Sp89,MSSh89,HY98}, $\ve$ algorithm of Wynn, "AW" \cite{M89}),
or found experimentally \cite{Mit86,Th88} and through the MC simulations
\cite{MC98}. To compare those
results with our numbers, the fixed point locations are given by taking the
normalizing factor $\frac{8}{9}$ into consideration.
As experimental data we consider the averaged values of Ref. \cite{Mit86}
obtained in a course of studying the critical behavior of the site-random
Ising system $Mn_x Zn_{1-x} F_2$ with $x=0.75$ (or 0.50) by the neutron
scattering method as well as the averaged values of Ref. \cite{Th88}.

In Table IV we present also the numerical results for the fixed point
locations and critical exponents obtained on the basis of the simple Borel
summation method combined with the Pad$\acute{\rm e}$ approximants
[3/1] and [2/2]. Although the fixed point coordinates obtained
in such a way turned out to be strongly different, the critical exponents
estimates differ nevertheless from each other only slightly. From this point
of view neither approximant is better.

The another possible way to determine the coordinates of the RIM fixed point
is to calculate them as averages between the values given by the highest
approximants [3/1] and [2/2] for each of the components (see Fig. 7).
The corresponding critical exponents estimates are found to agree with just
considered as well as with experimental data and MC results,
within the error bounds (see Table IV).

Unlike the cubic fixed point, the determination of the error bounds
for the RIM fixed point locations is a more difficult problem. In the case
of selecting the most stable approximants we have taken the following scheme.
First, the values of the fixed point locations given by the approximants
[3/1] and [2/2] are averaged for each of the components separately within the
interval $0 \le b \le 4$. The discrepancy between averages in
these approximants is then adopted as a sought uncertainty in the results.
At the same time,
the error bounds for the critical exponents are determined in the same way as
in Sec. V. If we adopt as the fixed point locations the averages between
the values given by the highest approximants [3/1] and [2/2] separately
for each of the components, the error bounds seem to be even smaller, they are
almost three quarters of the previous ones. Absence of the error
bounds in some of the places of Table IV means that the errors either
cannot be established or were not established.

We have checked also the stability of the RIM fixed point on the
three-dimensional RG flow diagram. In all considered cases the
eigenvalues of the stability matrix turned out to be negative,
except the calculations based on the simple Borel summation method with the
Pad$\acute{\rm e}$ approximant [2/2]. In this case the second eigenvalue
$\la_2$ occurred to be positive and too large. This is in contradiction
to the known theoretical and experimental predictions as well as to the
Monte-Carlo simulations \cite{MC98}. Indeed, the second, smallest in modulos,
eigenvalue $\la_2$ of the matrix of derivatives of the $\bt$-functions is
well known to define the so called correction-to-scaling exponent
$\om$ that governs the leading corrections to the universal
power laws. Thus, the approach of the zero field susceptibility to the
critical temperature, for $T > T_c$, is characterized by the Wegner
series \cite{Wr72}
\be
\chi \simeq \Gm_0 \tau^{- \gm} \bigl(1 + \Gm_1 \tau^{\om \nu}
   + \dots \bigr) ,
\label{eq:21}
\ee
with $\Gm_k$ being the nonuniversal amplitudes and $\tau=(T-T_c)/T_c$
\cite{BLGZ}. Here $\gm$, $\nu$ are the asymptotic values of the
susceptibility and correlation length critical exponents.
As the exponent $\om$ decreases the region increases, where
the corrections to scaling laws should be taken into account.
So, the smallness of $\om$ in the RIM indicates the importance of its
calculation for analysis of the asymptotic critical behavior of dilute
systems \cite{FHY98p}.

Recent MC calculations based on the analysis of the first correction term
in Eq.(\ref{eq:21}) provided the estimate of the correction-to-scaling
exponent $\om=0.37 \pm 0.06$ \cite{MC98}. Almost the same number was
more recently obtained in the framework of the four-loop 3D RG analysis
used for processing
divergent series the Borel summation method in combination with the simple
rational Chisholm approximants like [M,M/1,1], $\om=0.372 \pm 0.005$
\cite{FHY98p}. Although the apparent accuracy of this estimate seems to be
highly overstated, the central value is in accordance with previous
estimates $\om= 0.366$ \cite{JOS} and $\om= 0.359$ \cite{Sp89},
$\om= 0.376$ \cite{HY98} derived within the three-loop approximation
in the framework of the minimal subtraction scheme and
the 3D RG respectively. Our estimate of $\om$ obtained on the basis of
the Borel summation method with the Pad$\acute{\rm e}$ approximant [3/1] is
close to the above mentioned. On the contrary, using the Pad$\acute{\rm e}$
approximant [2/2] in the Borel transformation leads to the unphysical result
for the correction-to-scaling exponent.

At the same time, applying the PBL resummation method to study the asymptotic
critical behavior of the systems with impurities results in the
correction-to-scaling exponent values which are different from those
predicted by either MC simulations or simple resummation procedures
(see Table IV).
It is the consequence of the fact that the four-loop approximation is
not enough to obtain reliable estimates of the RIM fixed point locations.
Note, however, that our estimates of $\om$ are within the
error interval found for the MC result. The critical exponents estimates
obtained confirm also the inequality $- \nu \om < \al < 0$
conjectured for the random models \cite{AHW98}. Unfortunately, at present
we cannot indicate any error bounds in our calculation of the exponent
$\om$.

So, if to assume that the MC simulations \cite{MC98} provide the numerical
estimate of $\om$ which is close to the "exact" one, a question is to be
put forward: can the estimation of the correction-to-scaling exponent of the
dilute systems be used as an additional criterion of selection of the
resummation techniques to be employed? Probably the answer will be
given in the course of the further investigation of the critical properties
of the RIM within the higher-order RG approximations provided a more
sophisticated method of the series summation will be used, on a level
with the simple techniques.
Thus more recently, a new approach to summation of divergent field-theoretical
series has been suggested \cite{MV98e}. The method, based on the Borel
transformation combined with a conformal mapping, relies upon the stability
of the result of processing on the transformation parameters and therefore
does not require knowing the exact asymptotic behavior of the series. This
method has been tested on the functions expanded in their asymptotic power
series and applied to estimating the ground state energy of quantum mechanical
systems, including anisotropic oscillator, as well as to calculating
the critical exponents for some conformal field theories \cite{MV98e,MV99d}.
The successful testing the developed technique on simple systems made it
possible to apply it for obtaining the critical exponents estimates of
anisotropic $N$-vector field models describing both magnetic and structural
phase transitions in cubic and tetragonal crystals \cite{MV98-1,MV98e}.
It would be
reasonable therefore to apply this technique in studying the critical
behavior of the RIM exactly in three dimensions within the five-or
higher-loop approximations.

\section*{Conclusion}

The complete RG analysis of a field model with two quartic coupling constants
associated with isotropic and cubic interactions describing magnetic and
structural phase transitions in a good number of real substances has been
carried out within the four-loop approximation directly in three dimensions.
Perturbative expansions for $\bt$-functions and critical exponents were
deduced for generic $N$. The fixed points locations were found for $N\ge3$
by applying the generalized Pad$\acute{\rm e}$-Borel-Leroy resummation
technique, and the global structure of the 3D RG flow diagram was
investigated.
The analysis of the eigenvalue exponents of the most intriguing isotropic
and cubic fixed points fulfilled for the physically important case $N=3$ has
shown that the cubic rather than the isotropic fixed point is absolutely
stable in 3D. The eigenvalues estimates of both fixed points were found
to agree well with those of recently calculated on the basis of
exploiting the five-loop $\ve$-expansions \cite{KTSf-b}.
The critical dimensionality $N_c$ of the order parameter field,
at which the topology of the flow diagram changes, has been analyzed by
the two different methods:
1) by resumming the four-loop RG expansions for the $\bt$-functions
in 3D and 2) by resumming the five-loop $\ve$-expansion for $N_c$ at $\ve=1$.
The numerical estimates $N_c=2.89 \pm 0.02$ and $N_c=2.894 \pm 0.040$
obtained are in a good agreement with the earlier results \cite{Sp89,MSSh89}
and confirm the conclusion about the stability of the cubic fixed point for
$N\ge3$. Consequently, the magnetic and structural phase transitions in
three-dimensional anisotropic crystals with cubic symmetry are of second order
and their critical thermodynamics should be governed by the cubic fixed point
with a specific set of critical exponents. The corresponding four-loop
critical exponents estimates were found in the framework of the PBL
resummation method. On the basis of comparative numerical analysis with the
$O(N)$-symmetric model, the critical exponents of which are solid established
\cite{Z}, it has been shown that in the case of the cubic model the
difference between the four-loop estimates and the exact values does not
exceed 1--2\%. Although our results for the most interesting case $N=3$ are in
good accordance with earlier theoretical predictions, the cubic universality
class is not easy to distinguish experimentally from the isotropic one,
due to the obvious marginality of the problem, $N_c \sim 3$.

The critical behavior of weakly quenched disordered systems undergoing
second order phase transitions and described by the three-dimensional random
Ising model, which is the nontrivial specific case of the cubic model when
$N=0$, has also been investigated. The coordinates, eigenvalues and critical
exponents of the RIM fixed point were computed by using the PBL resummation
method and the more simple Pad$\acute{\rm e}$-Borel procedure. Our numerical
results along with the known theoretical and experimental data were
summarized in the table. Although the RIM fixed point locations found
in the framework of different approximation schemes turned out to be
strongly different, the critical exponents estimates differ from each other
only slightly, within the error bounds obtained.

A special attention was given to study the stability of the 3D RIM fixed
point
on the RG flow diagram. The calculation of the stability matrix eigenvalues
based on applying different resummation techniques showed that they are
negative. Consequently, the RIM fixed pont is stable in 3D. As an exception
we have indicated the case of using the simple Borel summation method in
combination with the Pad$\acute{\rm e}$ approximant [2/2], which led to the
unphysical result for the second eigenvalue $\la_2 = \om$.
Applying the PBL procedure to calculate the leading correction-to-scaling
exponent $\om$ for
the three-dimensional impure systems was shown to result in the numerical
estimates which are distinguishable markedly from those predicted by recent
MC simulations \cite{MC98} or followed from applying simple Borel-like
resummation procedures \cite{Sp89,HY98,FHY98p} (see Table IV). Note, however,
that our estimates of $\om$ were found to be within the error bounds
known for the MC result.

At last, it is worth noting that one can hardly hope to extract a reliable
numerical estimate of $\om$ from the five-loop $\ve$-expansion
\cite{KS95}, even resummed by a proper method.
The point is that, the $\ve$-series for $\om$ turns out
to be very short, due to the degeneracy of the random Ising model
$\bt$-functions on the one-loop level. Therefore, until the appreciable
discrepancy between the results given by different
methods of the series summation on the one hand, and the MC calculations,
on the other, does exist, the further investigation of the asymptotic
critical behavior of dilute systems will be highly desirable in the framework
of the higher-order (five-or six-loop) RG approximations provided a more
sophisticated resummation procedure, for instance the Borel transformation
combined with a conformal mapping, will be applied.

{\it Note added to manuscript.} More recently the six-loop study of
critical behavior of the 3D cubic model appeared \cite{CPV99}. Perturbative
expansions for $\bt$-functions and critical exponents deduced within
the massive field theory in fixed dimension have been resumed by means
of the Borel transformation combined with a conformal mapping that takes
into account the singularities of the Borel transform. The fixed point
locations, stability matrix eigenvalues and critical exponents estimates
obtained turned out to be essentially the same results as those of the
present work. In Appendix we present our numerical estimates for the cubic
model obtained on the basis of the six-loop expansions of Ref. \cite{CPV99}
when the PBL resummation procedure is applied. The critical behavior of 
three-dimensional random Ising systems has recently been studied within the 
five- and six-loop RG approximations in Ref. \cite{PS99} and Ref. \cite{PV00} 
respectively.

\section*{Acknowledgments}

It is pleasure to thank Drs. A. I. Mudrov, B. N. Shalaev and
Prof. A. I. Sokolov for a reading of the manuscript. I am especially
grateful to Dr. B. N. Shalaev for many illuminating discussions on
properties of the phase transitions of disordered systems. I thank
Dr. E. Vicari for useful correspondence and comments.

\section*{Appendix}

In this section we present our numerical estimates for the cubic fixed
point locations and critical exponents for some values of $N$ (see Table V)
using the six-loop expansions for RG functions recently obtained by
J. M. Carmona, A. Pelissetto, and E. Vicari \cite{CPV99}. We apply the
generalized PBL resummation technique.
The susceptibility and correlation length critical exponents
are estimated through the original series for $\gm^{-1}$ and $\nu$,
whereas the critical exponent $\et$ is obtained by the known scaling
relation: $\et = 2 - \gm / \nu$. It is seen that our estimates are in
excellent agreement with those of Ref. \cite{CPV99}, where the Borel
resummation procedure with a conformal mapping, that takes into account
the singularities of the Borel transform, has been applied.

\nid

\newpage
\begin{center}
{\large \bf Figure Captions}
\end{center}

\nid
Fig. 1. Curves demonstrating dependence of the results of calculating
the cubic fixed point locations a) $u_c$-component and b) $v_c$-component
on transformation parameter $b$ from the four-loop approximation for
$N = 3$. The upper curve ($\Diamond$) corresponds to the [2/1]
Pad$\acute{\rm e}$ approximant (three-loop approximation), while the middle
($\triangle$) and lower ($\Box$) curves correspond to the [2/2] and [3/1]
Pad$\acute{\rm e}$ approximants (four-loop approximation) respectively.

\vspace{0.5cm}

\nid
Fig. 2. The results of computation of the O(3)-symmetric fixed point
locations from the three- to the six-loop approximations obtained
on the basis of the PBL resummation method with eight types of the
approximants: [2/1] - $\Diamond$,
[3/1] - $\Box$, [2/2] - $\triangle$, [4/1] - full $\Diamond$,
[3/2] - full $\triangle$, [5/1] - full $\circ$, [4/2] - $\circ$,
[3/3] - $\times$. On the six-loop level under the optimal value of the
transformation parameter $b=4.5$ we obtain the estimate $g_c=1.392$.

\vspace{0.5cm}

\nid
Fig. 3. Graphs of dependence of the results of processing the series
a) $-\frac{\partial\bt_u}{\partial u}$,
b) $-\frac{\partial\bt_v}{\partial v}$,
c) $-\frac{\partial\bt_u}{\partial v}$, and
d) $-\frac{\partial\bt_v}{\partial u}$
on the parameter $b$ for $N=3$. The curves on the pictures a) and b)
are given in the same notations as in the previous figures, while for
the curves corresponding to the approximants [1/1] and [2/1] on the
pictures c) and d) the notations $\Diamond$ and $\Box$ are used
respectively.

\vspace{0.5cm}

\nid
Fig. 4. Dependence of the results of processing the $\ve$-series for
the critical dimensionality $N_c$ on transformation parameter
$b$. The crossing of the curves corresponding to the three Pad$\acute{\rm e}$
approximants gives the optimal value of the parameter $b$ at which
the estimate $N_c=2.894$ is obtained.

\vspace{0.5cm}

\nid
Fig. 5. Curves demonstrating dependence of the result of processing
susceptibility exponent $\gm$ of the cubic model on the parameter $b$
for $N=3$ in the four-loop approximation.

\vspace{0.5cm}

\nid
Fig. 6. Graphs of dependence of the results of processing the 3D RG
series for susceptibility exponent $\gm$ of the $O(3)$-symmetric model on
the parameter $b$ for the eight types of Pad$\acute{\rm e}$ approximants
in the framework of the PBL procedure.

\vspace{0.5cm}

\nid
Fig. 7. Graphs of dependence of the results of calculating
the RIM fixed point locations a) $u_c$-component and b) $v_c$-component
on transformation parameter $b$ within the four-loop approximation.
The curves marked by $\Diamond$, $\triangle$, and $\Box$ correspond to
the Pad$\acute{\rm e}$ approximants [1/1], [2/2], and [3/1] respectively.

\newpage
\begin{table}[1]
\caption{Coordinates of the cubic fixed point of RG equations
 for $N=3$ found under the optimal value of the transformation
 parameter $b=0$. "AV" denotes the average value.}
\vspace{0.5cm}
\hspace{0.2cm}
\begin{tabular}{c|c|c|c|c|c|c|c}
\hline
\hline
& [2/1] & [3/1] & [2/2] & AV & AV over [2/2] & Ref. \cite{Sp89} &
Ref. \cite{MSSh89} \\
\hline
\hline
$u_c$ & 1.3536 & 1.3338 & 1.3410 & 1.3428 & 1.3425 & 1.348 & 1.3357 \\
      &        &        &        & $\pm$ 0.0200 &  &       & \\
\hline
$v_c$ & 0.0526 & 0.1026 & 0.0894 & 0.0815 & 0.0937 & 0.090 & 0.0906 \\
      &        &        &        & $\pm$ 0.0300 &  &       & \\
\hline
\hline
\end{tabular}
\label{table 1}
\end{table}

\newpage
\begin{table}[2]
\caption{Coordinates of the cubic fixed point of RG equations and
 critical exponents estimates for some $N$ found under the optimal
 value of the transformation parameter $b$ within the four-loop
 approximation. The critical exponents values obtained within the
 framework of the $\ve$-expansion method (five-loop results) and
 marked by the symbol (*) are presented for comparison.}
\vspace{0.5cm}
\hspace{2.5cm}
\begin{tabular}{c|c|c|c|c|c}
\hline
\hline
N  & $u_c$ & $v_c$  & $\et$  & $\nu$  & $\gm$  \\
   &       &        &        &        &        \\
\hline
\hline
3 & 1.3428 & 0.0815 & 0.0332 & 0.6996 & 1.3775 \\
  &        &        & 0.0375$^*$ & 0.6997$^*$ & 1.3746$^*$ \\
\hline
4 & 0.9055 & 0.8167 & 0.0327 & 0.7131 & 1.4028 \\
  &        &        & 0.0365$^*$ & 0.7225$^*$ & 1.4208$^*$ \\
\hline
5 & 0.6980 & 1.2361 & 0.0325 & 0.7154 & 1.4076 \\
  &        &        & 0.0358$^*$ & 0.7290$^*$ & 1.4305$^*$ \\
\hline
6 & 0.5807 & 1.5386 & 0.0324 & 0.7157 & 1.4082 \\
  &        &        & 0.0354$^*$ & 0.7301$^*$ & 1.4322$^*$ \\
\hline
7 & 0.5060 & 1.7874 & 0.0324 & 0.7155 & 1.4079 \\
\hline
8 & 0.4544 & 2.0076 & 0.0324 & 0.7153 & 1.4074 \\
\hline
9 & 0.4168 & 2.2108 & 0.0324 & 0.7149 & 1.4067 \\
\hline
10 & 0.3881 & 2.4032 & 0.0324 & 0.7147 & 1.4063 \\
\hline
12 & 0.3473 & 2.7680 & 0.0323 & 0.7142 & 1.4054 \\
\hline
\hline
\multicolumn{5}{l}{\footnotesize $^{*}$ Quoted from
Ref. \cite{MV98e}}
\end{tabular}
\label{table 2}
\end{table}

\newpage
\begin{table}[3]
\caption{Four-loop eigenvalue exponents estimates for the cubic (CFP)
 and Heisenberg (HFP) fixed points ($N=3$) found in $3D$ under the
 optimal value of the transformation parameter $b$.}
\vspace{0.5cm}
\hspace{2cm}
\begin{tabular}{c|c|c|c|c}
\hline
\hline
& CFP   & CFP, Ref. \cite{KTSf-b} & HFP & HFP, Ref. \cite{KTSf-b}  \\
\hline
\hline
$\la_1$ & -0.7786  & -0.7648  & -0.7791 & -0.7640 \\
\hline
$\la_2$ & -0.0081  & -0.0085  & 0.0077  &  0.0089 \\
\hline
\hline
\end{tabular}
\label{table 3}
\end{table}

\begin{table}[4]
\caption{Numerical results for the RIM. Here "MSA" indicates that
 the fixed point locations and the critical exponents are
 found for the most stable approximants in the framework of the PBL
 resummation technique, "AV" means that the fixed point locations
 are calculated as the averages between the values given by the highest
 approximants [3/1] and [2/2], "PB" denotes a simple Pad$\acute{\rm e}$-Borel
 resummation. The results obtained either within the different theoretical
 approaches, or on the basis of using different resummation techniques,
 or experimentally ("Exp.") and by means of the MC simulations, are
 presented for comparison.}
\vspace{0.5cm}
\begin{tabular}{c|c|c|c|c|c|c}
\hline
\hline
           & $u_c$ & $v_c$  & $\et$  & $\nu$  & $\gm$  & $\om$   \\
           &       &        &        &        &        &         \\
\hline
\hline
MSA        & -0.5816 & 1.9822 & 0.040 & 0.681 & 1.336 & 0.310 \\
           & $\pm$ 0.0850 & $\pm$ 0.0740 & $\pm$ 0.011 & $\pm$ 0.012
           & $\pm$ 0.020 & \\
\hline
AV         & -0.6246 & 1.9438 & 0.034 & 0.672 & 1.323 & 0.330 \\
           & $\pm$ 0.0600 & $\pm$ 0.0500 & $\pm$ 0.010 & $\pm$ 0.004
           & $\pm$ 0.010 & \\
\hline
PB [3/1]   & -0.6839 & 1.9877 & 0.033 & 0.674 & 1.326 & 0.362  \\
\hline
PB [2/2]   & -0.5800 & 1.8934 & 0.034 & 0.669 & 1.316 & \\
\hline
AW         & -0.5874$^a$ & 1.9362$^a$ &  & 0.668$^a$ & 1.318$^a$ & \\
\hline
Ref. \cite{Sp89} & -0.728 & 2.006 & 0.021 & 0.671 & 1.328 & 0.359  \\
\hline
Ref. \cite{HY98} & -0.745 & 2.011 & 0.019 & 0.671 & 1.328 & 0.376  \\
\hline
Ref. \cite{MSSh89} & -0.6668 & 1.9951 & 0.034 & 0.670 & 1.326 &    \\
\hline
3D MS             &  &  & 0.053$^b$ & 0.677$^b$ & 1.319$^b$ & 0.330$^b$ \\
                  &  &  &           &           &           & 0.390$^e$ \\
\hline
Exp.       &  &  &  & 0.71 $\pm$ 0.02$^c$ & 1.37 $\pm$ 0.04$^c$ & \\
           &  &  &  & 0.70 $\pm$ 0.02$^{a,d}$ & 1.37 $\pm$ 0.04$^{a,d}$ & \\
\hline
MC,        &  &  &  & 0.6837 & 1.342 & 0.37 \\
Ref. \cite{MC98} &  &  &  & $\pm$ 0.0053 & $\pm$ 0.010 & $\pm$ 0.06 \\
\hline
\hline
\multicolumn{5}{l}{\footnotesize $^a$ Quoted from
Ref. \cite{M89}} \\
\multicolumn{5}{l}{\footnotesize $^b$ Quoted from
Ref. \cite{FHY98}} \\
\multicolumn{5}{l}{\footnotesize $^c$ Quoted from
Ref. \cite{Mit86}} \\
\multicolumn{5}{l}{\footnotesize $^d$ Quoted from
Ref. \cite{Th88}} \\
\multicolumn{5}{l}{\footnotesize $^e$ Quoted from
Ref. \cite{FHY98p}}
\end{tabular}
\label{table 4}
\end{table}

\begin{table}[5]
\caption{The cubic fixed point locations and critical exponents
 estimates for some $N$ obtained within the PBL resummation procedure
 in the six-loop approximation. Here the coupling constants are given
 in the same notations as those of Ref. \cite{CPV99}.}
\vspace{0.5cm}
\hspace{0.2cm}
\begin{tabular}{c|c|c|c|c|c|c|c}
\hline
\hline
N  & $u_c$       & $v_c$        & $\et$  & $\nu$  & $\gm$  &
$\la_1$ \cite{N} & $\la_2$ \cite{N} \\
   &             &              &        &        &        &
         &          \\
\hline
\hline
3 &    1.3177    &    0.0964    & 0.0327 & 0.7040 & 1.3850 &
     - 0.7833    &  - 0.0109    \\
  & $\pm$ 0.0170 & $\pm$ 0.0165 & $\pm$ 0.0020 & $\pm$ 0.0040 & $\pm$ 0.0050 &
    $\pm$ 0.0054 & $\pm$ 0.0032 \\
\hline
4 &    0.8804    &    0.6360    & 0.0316 & 0.7150 & 1.4074 &
     - 0.7887    &  - 0.0740    \\
  & $\pm$ 0.0080 & $\pm$ 0.0050 & $\pm$ 0.0025 & $\pm$ 0.0050 & $\pm$ 0.0030 &
    $\pm$ 0.0090 & $\pm$ 0.0065 \\
\hline
8 &    0.4410    &    1.1331    & 0.0305 & 0.7143 & 1.4068 &
     - 0.7955    &  - 0.1396    \\
  & $\pm$ 0.0070 & $\pm$ 0.0160 & $\pm$ 0.0025 & $\pm$ 0.0035 & $\pm$ 0.0030 &
    $\pm$ 0.0150 & $\pm$ 0.0100 \\
\hline
$\infty$ &    0.1751    &    1.4122    & 0.0319 & 0.7094 & 1.3962 &
            - 0.7986    &  - 0.1787   \\
  & $\pm$ 0.0040 & $\pm$ 0.0090 & $\pm$ 0.0035 & $\pm$ 0.0030 & $\pm$ 0.0040 &
    $\pm$ 0.0200 & $\pm$ 0.0050 \\
\hline
\hline
\end{tabular}
\label{table 5}
\end{table}


\begin{thebibliography}{99}
\bibitem{13} It was established that the diagonal or near-diagonal
Pad$\acute{\rm e}$ approximants exhibit the best approximating properties.
See, for instance, \cite{BGm}
\bibitem{N} The stability matrix eigenvalues estimates obtained in the
framework of the PBL resummation technique with the use of approximants
[5/1] and [4/1] for the series $\frac{\partial\bt_u}{\partial u}$,
$\frac{\partial\bt_v}{\partial v}$ and approximants
[4/1] and [3/1] for the series $\frac{\partial\bt_u}{\partial v}$,
$\frac{\partial\bt_v}{\partial u}$ under the optimal value of the
transformation parameter $b$ are $\la_1 = -0.7887$ and
$\la_2 = -0.0126$ for $N=3$, $\la_1 = -0.7895$ and
$\la_2 = -0.0760$ for $N=4$, $\la_1 = -0.8004$ and
$\la_2 = -0.1414$ for $N=8$, $\la_1 = -0.8054$ and
$\la_2 = -0.1769$ for $N=\infty$.
\end{thebibliography}
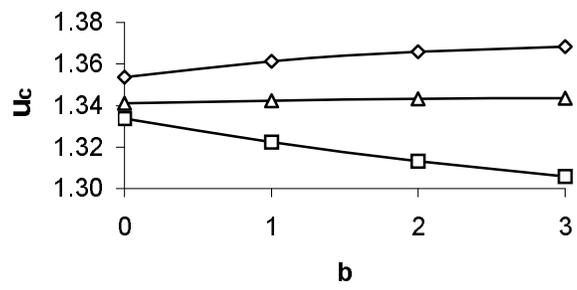
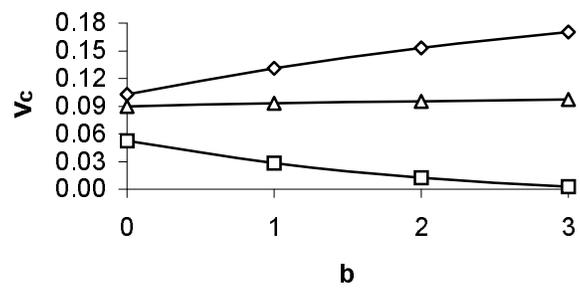
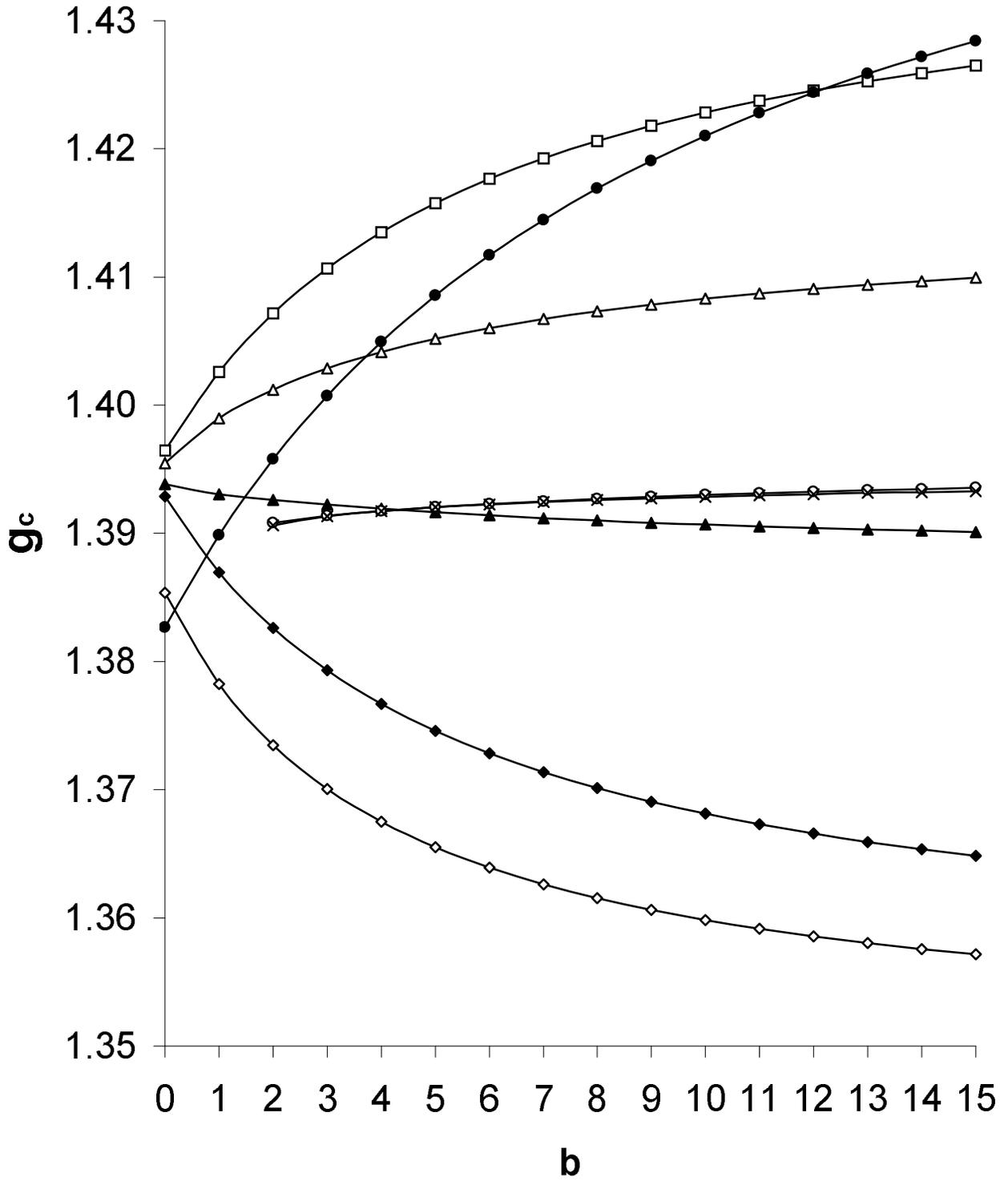
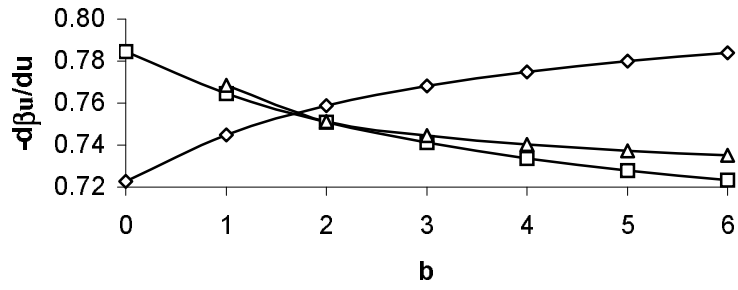
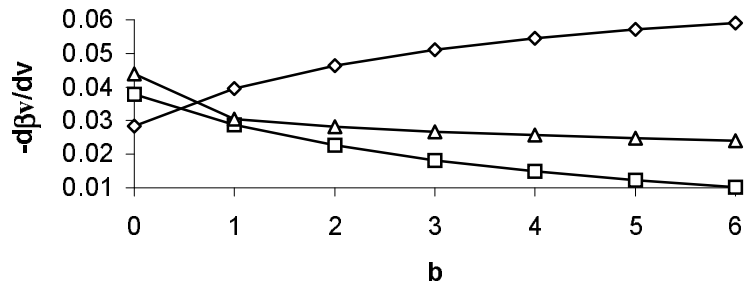
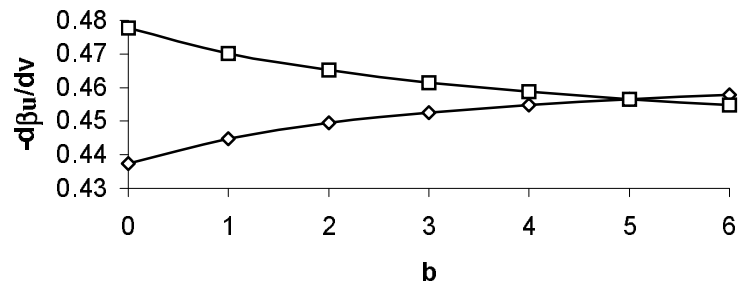
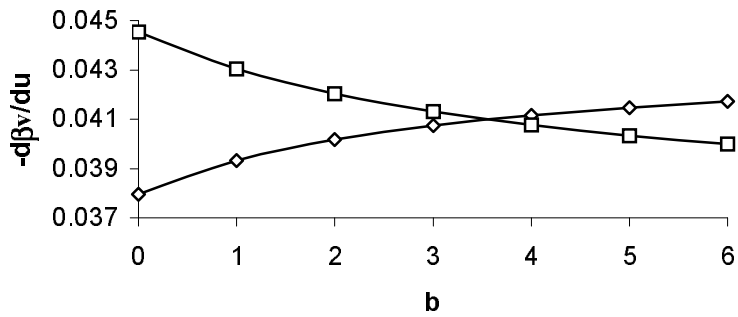
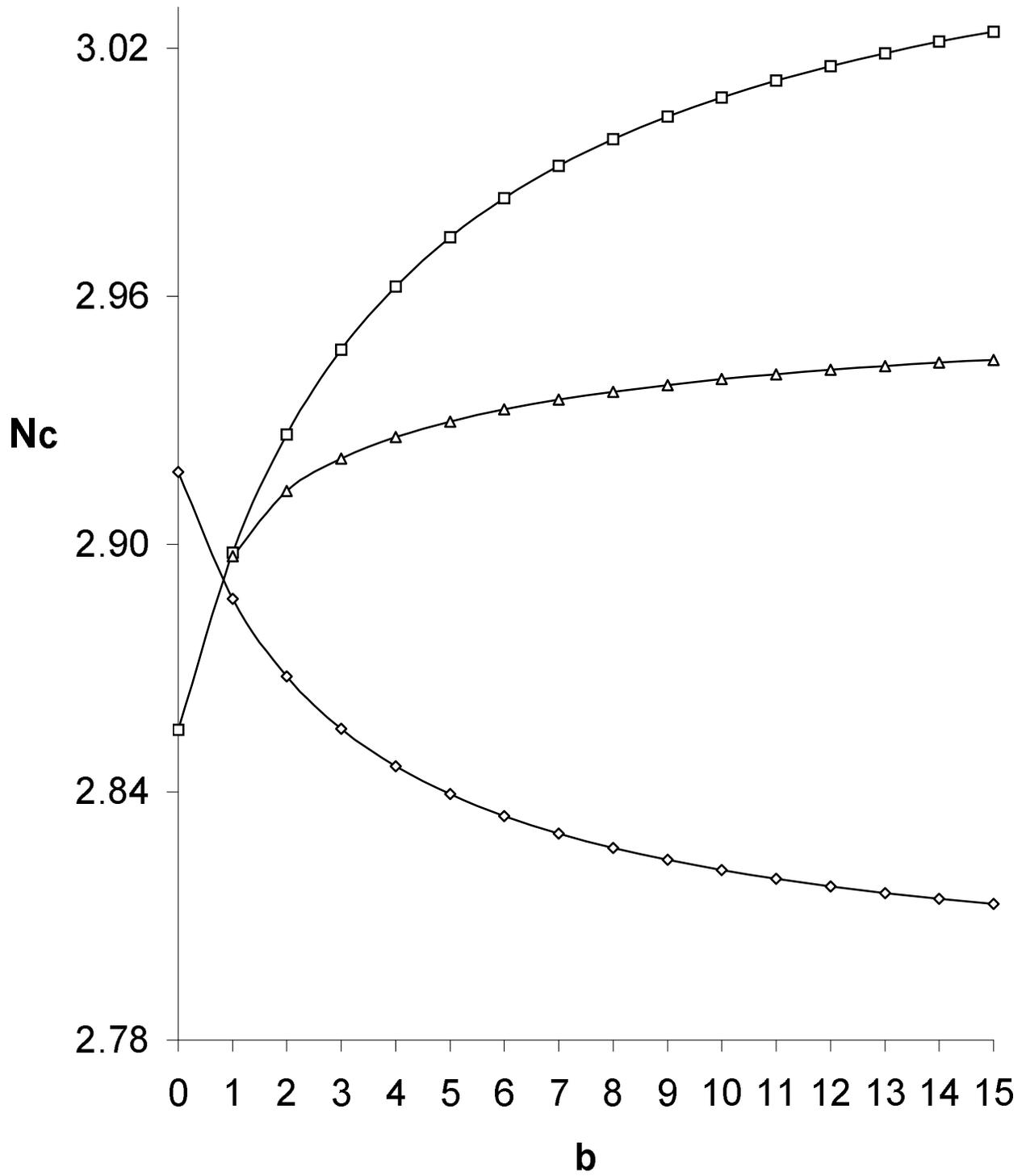
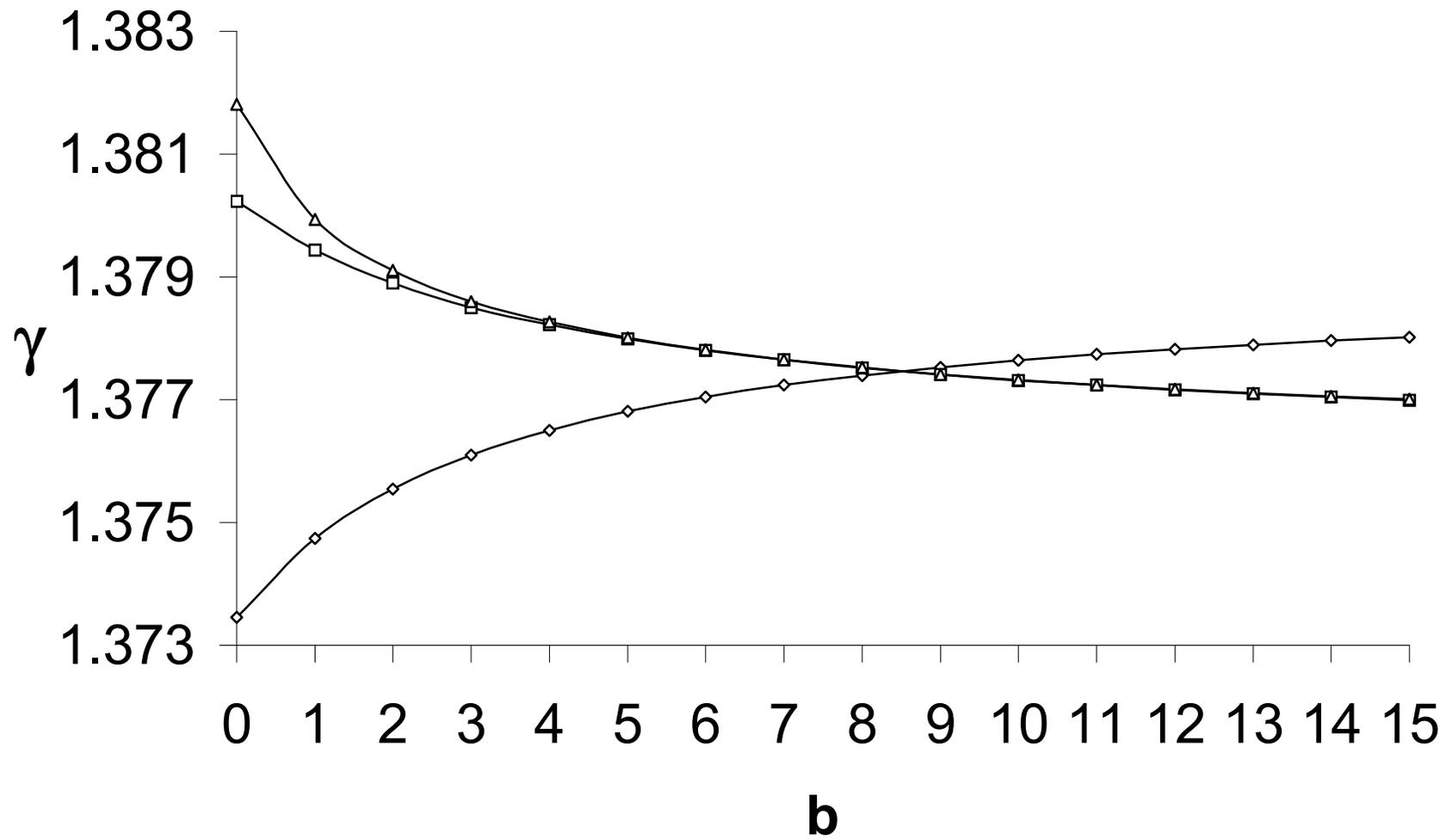
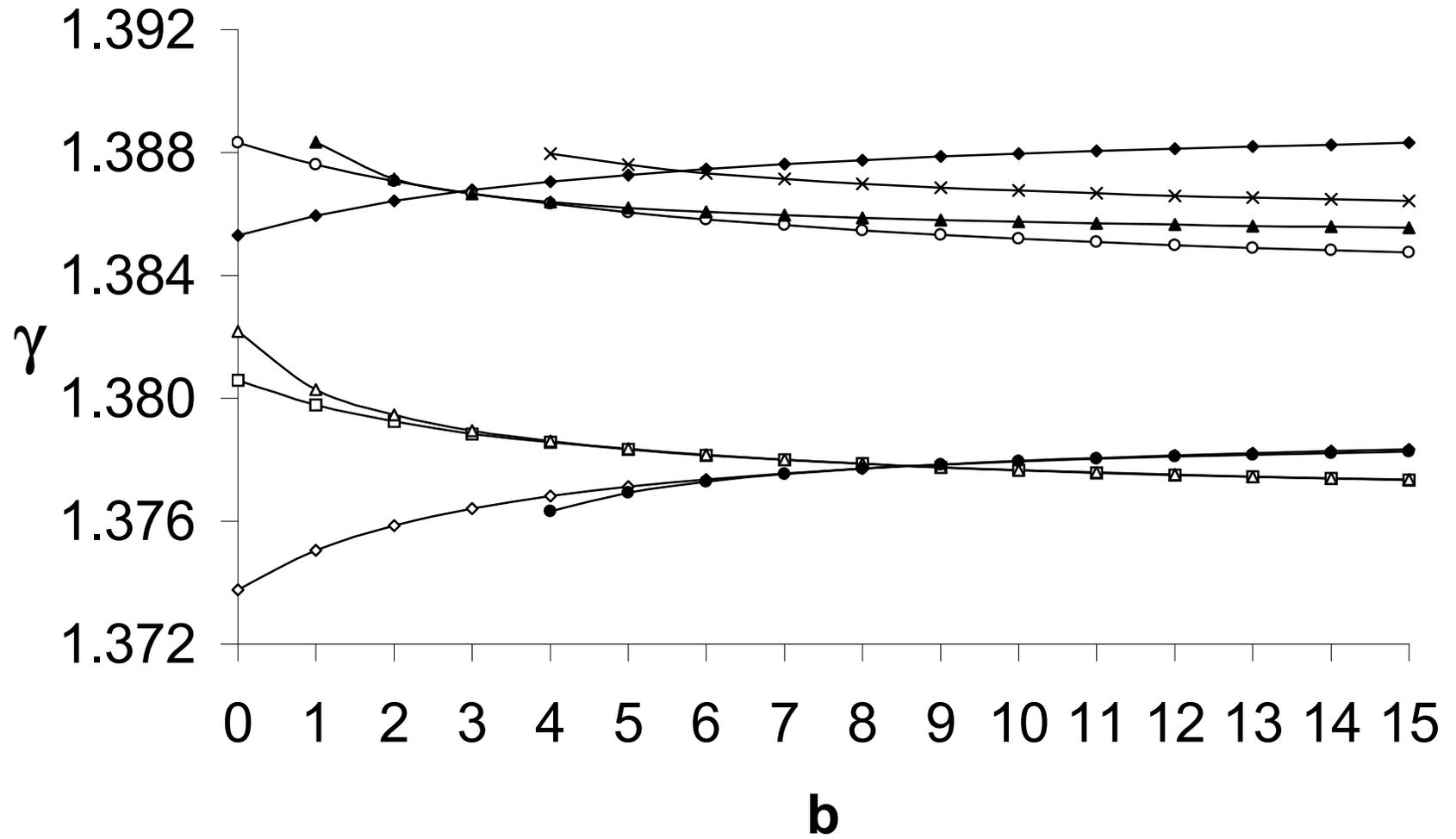
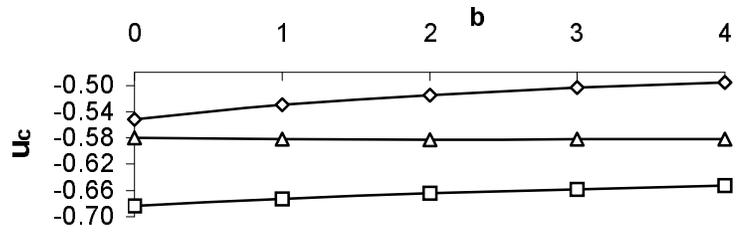
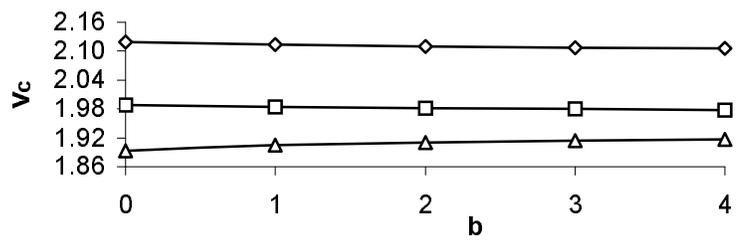
\end{document}